# A Slot Antenna Array with Reconfigurable RCS Using Liquid Absorber

Yukun Zou, Xiangkun Kong, *Member*, *IEEE*, Lei Xing, *Member*, *IEEE,* Shunliu Jiang, Xuemeng Wang, He Wang, Zhiming Liu, Yongjiu Zhao, Jens Bornemann, *Life Fellow, IEEE*

*Abstract*—**This paper presents a slot antenna array with a reconfigurable radar cross section (RCS). The antenna system is formed by combining a liquid absorber with a 2×2 slot antenna array. The liquid absorber consists of a polymethyl methacrylate (PMMA) container, a 45% ethanol layer, and a metal ground, which is attached to the surface of the slot antenna array. The incident wave can be absorbed by the absorber rather than reflected in other directions when the PMMA container is filled with ethanol, which reduces the monostatic and bistatic RCS. Thus the RCS of the antenna can be changed by injecting and extracting ethanol while the antenna's radiation performance in terms of bandwidth, radiation patterns and gain is well sustained. In a complex communication system, this can be used to switch between detection and stealth mode. The mechanism of the absorber is investigated. The simulated results show that the antenna with this absorber has monostatic and bistatic RCS reduction bands from 2.0 GHz to 18.0 GHz, a maximum RCS reduction of 35 dB with an average RCS reduction of 13.28 dB. The antenna's operating band is 100 MHz. Without ethanol, the antenna has a realized gain of 12.1 dBi, and it drops by 2 dB when the lossy ethanol is injected. The measured results agree well with the simulated ones.**

## INTRODUCTION

RADAR cross section (RCS) topics are of fundamental importance in defense electronics, and related antennas usually contribute to good RCS values for stealth platforms [1]. Different from the RCS reduction of common objects, it is

This work was supported in part by the National Natural Science Foundation of China under Grant 62071227, in part by the Natural Science Foundation of Jiangsu Province of China under Grant BK20201289, in part by the Open Research Program in China's State Key Laboratory of Millimeter Waves under Grant K202027 in part by the Postgraduate Research & Practice Innovation Program of Jiangsu Province under Grant SJCX20_0070, and in part by the Natural Science and Engineering Research Council (NSERC) of Canada.

Y. Zou, X. Kong, L. Xing, S. Jiang, X. Wang, H. Wang and Y. Zhao are with the College of Electronic and Information Engineering, Nanjing University of Aeronautics and Astronautics, Nanjing, 211106, China (e-mail: xkkong@nuaa.edu.cn).

Z. Liu is with the School of Information Engineering, Nanchang University, Nanchang 330031, China.

J. Bornemann is with the Department of Electrical and Computer Engineering, University of Victoria, Victoria, BC, V8W 2Y2, Canada.

necessary in this scenario to maintain the radiation characteristics of antennas while having low scattering properties.

Recently, several approaches have been proposed to reduce the RCS of antennas. In these techniques, metasurfaces were widely used, and metasurfaces with filtering features have been applied as excellent measures to reduce the RCS. For example, band-pass frequency selective surfaces (FSSs) as radomes [2], [3] or a band-notched FSS as a metal ground [4]-[5] were focused on reducing the out-of-band RCS of antennas. Another option is to load the antenna with metamaterial absorbers [6], [7]. In this way, the in-band and out-of-band RCS will be reduced but sometimes the antenna gain is compromised. Furthermore, chessboard layout matasurfaces were employed to reduce the antenna RCS through the phase cancellation between incident and reflected waves [8]-[11]. This method can effectively reduce the in-band RCS. Moreover, polarization conversion metasurfaces were used to reduce the RCS of antennas [12]-[16]. This technology can achieve ultra-wideband RCS reduction, but is only effective for the monostatic RCS. An integrating design of antennas and low-scattering metasurfaes can achieve low-RCS systems, while the radiation characteristics are well maintained [17]-[18]. The combination of Fabry-Perot (FP) resonator antennas and low-scattering metasurfuces is an excellent way to deal with gain reduction [19]-[22]. Specifically, combining a polarization conversion metasurface with a metamaterial absorber can achieve superior RCS reduction [23]. In this method, the in-band co-polarized waves will be transformed into their cross-polarized waves and then be absorbed, while the out-of-band waves will be absorbed directly.

The low RCS antennas discussed above fall in the category of passive devices, which are mostly inflexible after fabrication. Unfortunately, switching and multifunction antennas are frequently required in practical applications due to changes in a real environment. Nowadays, some liquid absorbers have been proposed because of their wide absorption band, transparency, reconfigurable properties, low-RCS, and low price of the liquid materials. A liquid material can be used as a radiator in one design [24]-[25], or it can be used as a unit cell to absorb incident waves in another one [26]-[28].

To overcome the challenge of RCS reconfigurability, this paper focuses on a three-layer liquid absorber to reduce the wideband RCS of a 2×2 slot antenna array. The liquid absorber can absorb incident waves, thereby reducing the RCS. The absorption band of the absorber ranges from 6.1 GHz to 20.54 GHz. The simulated results show that the antenna equipped with this absorber exhibits monostatic and bistatic RCS reduction bands from 2.0 GHz to 18.0 GHz, a maximum RCS reduction of 35 dB with an average RCS reduction of 13.28 dB. The reconfigurable property of the antenna is performed by injecting and extracting ethanol in a polymethyl methacrylate (PMMA) container. The mechanism and functionality of the liquid absorber is analyzed, and the RCS reconfigurable performance of the antenna is investigated.

This paper is organized as follows. Section II introduces the geometry and mechanism of the absorber. Section III presents the radiation parameters and scattering properties of the low-RCS slot array antenna with absorber, as well as the design and construction of the antenna. Section IV concludes the paper.

# I. DESIGN OF THE ABSORBER

## A. The permittivity of 45% ethanol

The liquid absorber consists of a PMMA container, a 45% ethanol (55% water) layer, and metal ground. The permittivity of the 45% ethanol at radio frequency can be represented using the Debye formula with the filled ethanol temperature of $T$ as follows [29]:

$$\varepsilon(\omega,T) = \varepsilon_\infty(T) + \frac{\varepsilon_0(T) - \varepsilon_\infty(T)}{1 - i\omega\tau(T)} \quad (1)$$

where $\varepsilon_\infty(\omega,T)$, $\varepsilon_0(T)$, $\tau(T)$ are the optical permittivity, static permittivity, and rotational relaxation time, respectively. They are only related to temperature and can be expressed as

$$\varepsilon_0(T) = a_1 - b_1T + c_1T^2 - d_1T^3 \quad (2)$$

$$\varepsilon_\infty(T) = \varepsilon_0(T) - a_2 e^{-b_2T} \quad (3)$$

$$\tau(T) = c_2 e^{\frac{d_2}{T+T_0}} \quad (4)$$

where $a_1$, $a_2$, $b_1$, $b_2$, $c_1$, $c_2$, $d_1$ and $d_2$ are constant variables of $\varepsilon_0(T)$, $\varepsilon_\infty(T)$ and $\tau(T)$, respectively. Here we set $T = 25$ ℃, so $\varepsilon(\omega,T)$ is only related to frequency. First, the permittivity of 45% ethanol is measured by Dielectric Assessment Kit V 2.4. Then the above equations are combined with the least square method to fit the permittivity of the 45% ethanol. The fitting curves are in good agreement with the measured ones, as shown in Fig. 1. It can be seen that the imaginary part of the permittivity is large, which ensures that the liquid absorber has good absorption property. The fitting curves of the permittivity will be applied to the full-wave electromagnetic simulation to obtain the features of the liquid absorber and the antenna array. The parameters are shown as follows: $\varepsilon_\infty(T) = 5.0344$, $\varepsilon_0(T) = 735.1702$, $\tau(T) = 20.9273$.

The reason for selecting 45% ethanol instead of all water is shown in Fig. 2. The real part of the permittivity of ethanol is substantially lower than that of water, thus enabling the liquid layer to be designed thicker, which assists us in making full use of the liquid's fluidity.

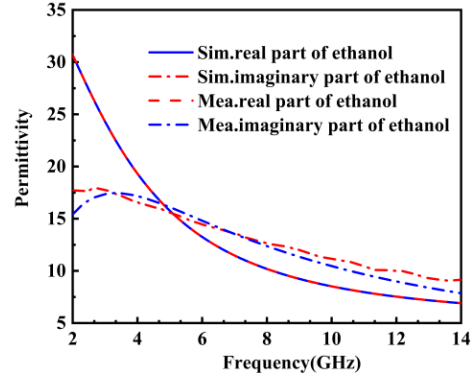

Fig. 1  Permittivity of 45% ethanol.

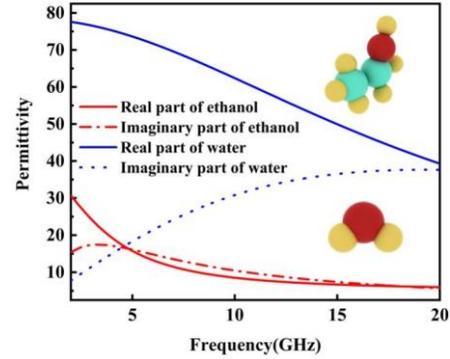

Fig. 2  Comparison of 45% ethanol and water.

## B. Structure of the Absorber

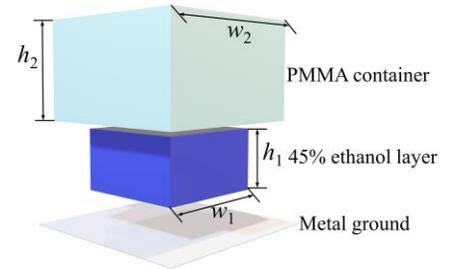

(a)

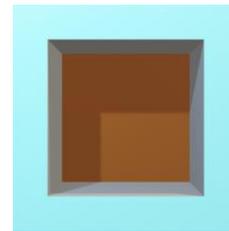

(b)

Fig. 3  (a) Unit cell structure of the absorber; (b) the bottom of the container.

The proposed liquid absorber is a sandwich structure composed of a container layer, a 45% ethanol layer and metal ground, as shown in Fig. 3(a). The container is made of PMMA ($\varepsilon_r$=2.67, tan$\delta$=0.01) on the top layer, with square grooves that match the shape of the square-type ethanol, as shown in Fig. 3(b). With the metal ground, the container produces an enclosed chamber that will be filled with ethanol. The length and width of the PMMA container are $w_2$=13 mm, and its thickness is $h_2$=9 mm The length and width of the square-type ethanol layer are $w_1$=5 mm, and its thickness is $h_1$=5 mm.

The $S_{11}$-parameter of the liquid absorber are investigated using full-wave electromagnetic simulation, and the results with different thickness $h_1$ are shown in Fig. 4. The absorption band ($S_{11}$ < 10 dB) ranges from 6.1 GHz to 20.54 GHz. The absorption band is mainly affected by the thickness of the ethanol layer $h_1$. The absorption band moves to higher frequencies with the increase of $h_1$ because the resonance length of the absorber becomes shorter. The absorption band will be narrow when $h_1$ changes due to the fact that the impedance of the absorber does not match that of free space in some frequency bands. Finally, in order to get good absorption property, $h_1$=5 mm is selected.

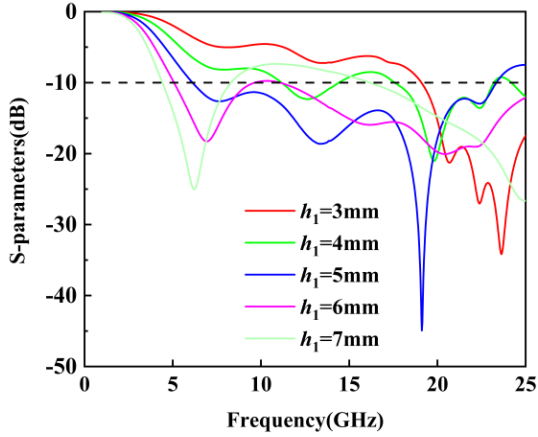

Fig. 4  Simulation results of the absorber.

## C.  Mechanism of the Absorber

To better understand the mechanism of the absorber, we use a one-port network, as proposed in [30], to describe the impedance and $S$-parameters of the liquid absorber. For a parallel resonant circuit, the real part of the impedance is maximum and the imaginary part is minmum. Therefore, we can obtain the impedance of the absorber from full-wave electromagnetic simulation from which we find that there are six maxima in the real part of the impedance. Thus six parallel resonant circuits with resonant frequencies $f_1$ to $f_6$, and unloaded $Q$ factors $Q_1$ to $Q_6$ are shown in Fig. 5. The series inductance and capacitance are described by $X_L(f)=X_{L0}f / f_c$ and $X_C(f)=X_{C0}f / f_c$, respectively, where $f_c$ is the center frequency. In the circuit model, resonant frequencies, unloaded $Q$ factors and $X_{L0}$ and $X_{C0}$ can be found by a genetic algorithm. Thus the impedance of the equivalent circuit model can be achieved and its $S$-parameters can be calculated by $Z=(1+|S_{11}|)/(1-|S_{11}|)$. The parameters are obtained as follows: $Q_1$=5.28, $Q_2$=4.38, $Q_3$=2.48, $Q_4$=2.52, $Q_5$=4.22, $Q_6$=4.01, $f_1$=4.28 GHz, $f_2$=4.72 GHz, $f_3$=5.68 GHz, $f_4$=11.39 GHz, $f_5$=15.5 GHz, $f_6$=20.95 GHz, $X_{L0}$=141.2 Ω, $X_{C0}$=1.86 Ω. The fitting results are shown in Fig. 6(a) and (b).

The curves of impedance and $S$-parameters in ADS are slightly different from those in full-wave electromagnetic simulation because the number of cascaded parallel resonant circuits is low. It is worth noting that the use of more cascaded parallel resonant circuits can achieve more accurate curves of the impedance and $S$-parameters.

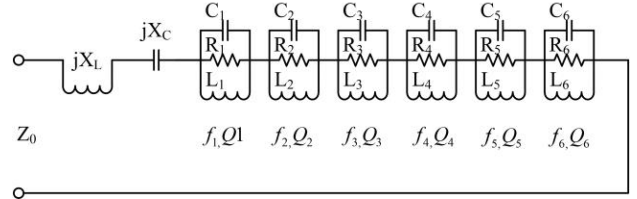

Fig. 5  Equivalent circuit model of the absorber.

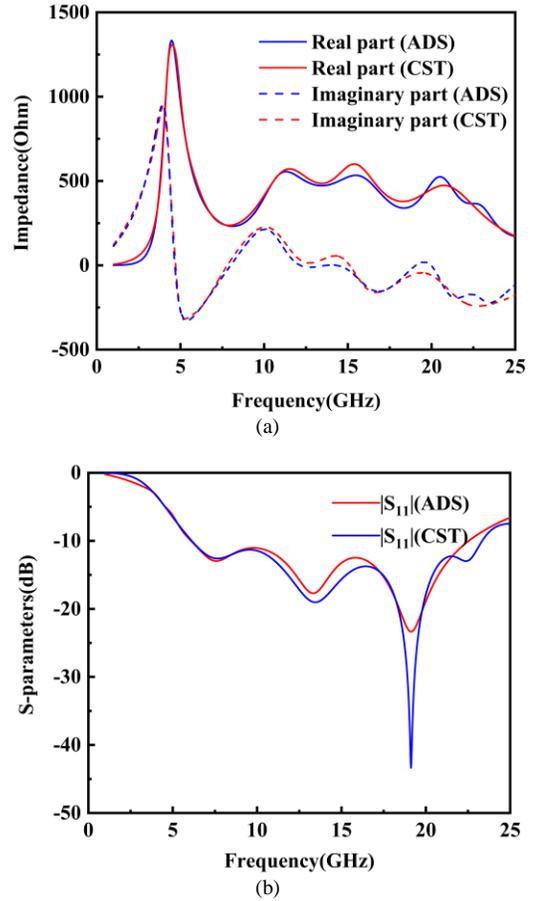

(a)

(b)

Fig. 6  (a) Comparison of impedance between ADS and CST; (b) comparison of s-parameters between ADS and CST.

The power-loss density of the absorption minima at 7.6 GHz, 13.3 GHz and 19.1 GHz are presented in Fig. 7 (a), (b) and (c). It is observed that at 7.6 GHz, the power is mainly concentrated on the top left and top right of the ethanol layer, the power is transfered to the upper surface of the ethanol layer at 13.3 GHz; and the majority of the power is localized at the interface of the ethanol layer and container layer at 19.1 GHz. This result reveals that the power is absorbed by the ethanol layer.

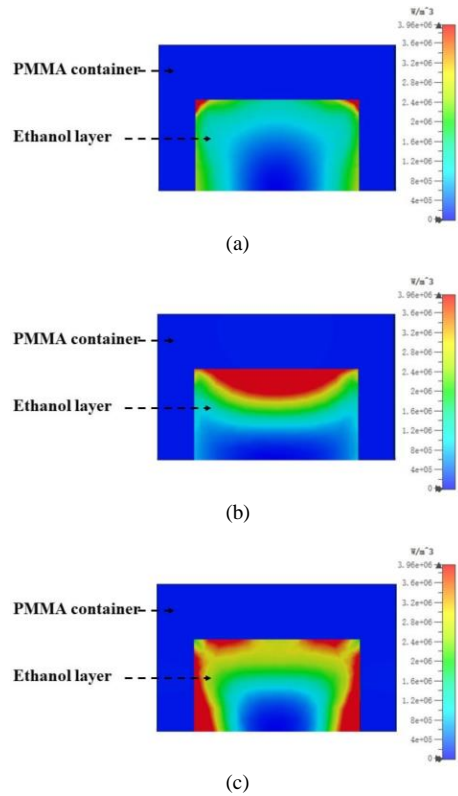

Fig. 7 Power-loss density. (a) $f$=7.6GHz, (b) $f$=13.3GHz, (c) $f$=19.1GHz.

## II. Design and Measurement of the Slot Antenna Array with Reconfigurable RCS

### A. Antenna Structure

To verify the reconfigurable low-RCS function of the liquid absorber, we combine the absorber with a 2×2 slot antenna array, as shown in Fig. 8(a). A metal ground with four slots is located on the top surface of a 1-mm thick FR4 dielectric substrate ($\varepsilon_r$=4.4, tan$\delta$=0.02). $W$ and $L$ represent the width and length of the metal ground, respectively. The width and length of the slot are $w$ and $l$, respectively, and the distance between slots is $d$. The feeding network is printed on the back of the FR4. A metal reflector is added to suppress the back lobe. The container with four slots is set on the surface of the metal ground which have been indicated in Fig. 8(a). The slots on the container are wider and longer than those on the metal ground, which can assist the slot antenna array to maintain its radiation performance. A square space is adopted to contain the ethanol, and a slot is designed between the adjacent square spaces to ensure the fluidity of ethanol, as shown in Fig. 8(b). The parameters of the antenna are: $W$=120 mm, $L$=220 mm, $w$=3 mm, $l$=33.2 mm, $d$=30 mm.

### B. Simulated and Measured Radiation Performance

The antenna has been fabricated and measured to demonstrate the reliability of our simulations.

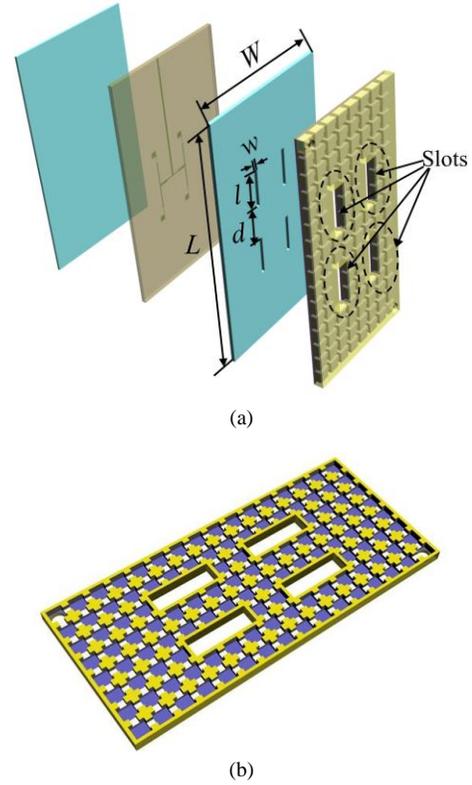

Fig.8  (a) 3-D sketch of the antenna; (b) back structure of the container.

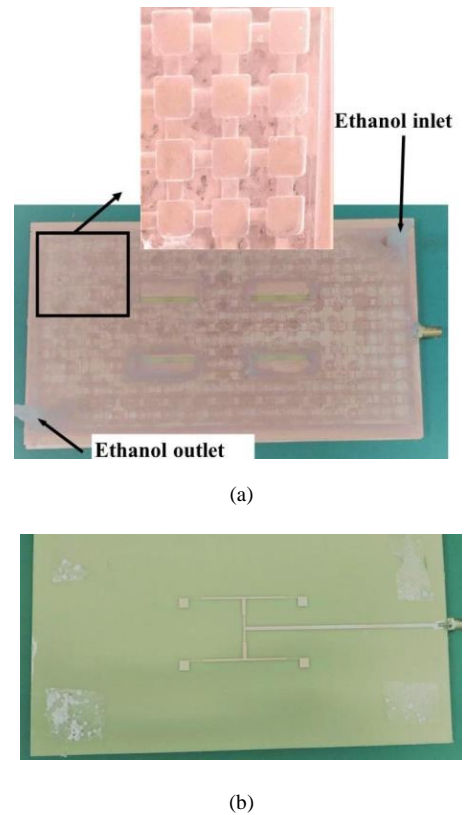

Fig. 9  Photographs of the fabricated (a) antenna and (b) feeding network of the antenna.

Fig. 9 (a) and (b) show photographs of the antenna and the feeding network. A 5245A vector network analyzer was used to measure the reflection coefficient, and the radiation patterns were measured in a microwave anechoic chamber, as shown in Fig. 10.

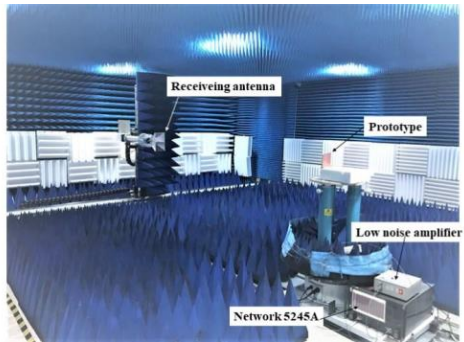

Fig. 10  The measurement setup of the radiation performance.

Fig. 11 (a) and (b) present the reflection coefficient of the antenna with ethanol or without ethanol. The 10-dB return-loss bandwidth of the antenna with or without ethanol is approximately 100 MHz, ranging from 2.94 GHz to 3.04 GHz. The difference between simulation and measurement is attributed to the glass glue used to seal the ethanol, which slightly changes the impedance of the antenna. The simulated and measured radiation patterns of the antenna with and without ethanol at 3 GHz are shown in Fig. 12 (a) and (b). The 3-D radiation patterns of the antenna without and with ethanol are depicted in Fig. 13 (a) and (b).

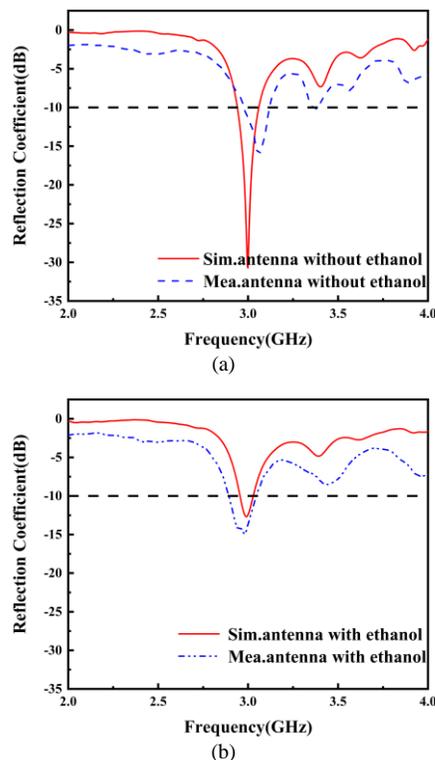

Fig. 11  Simulated and measured reflection coefficients of the antenna (a) without ethanol and (b) with ethanol.

The radiation patterns reveal that the radiation performance of the antenna without or with ethanol remains nearly identical, which coincides with our initial hypothesis. The realized gain of the antenna without ethanol is 12.1 dBi at 3 GHz. As the layer is filled with ethanol, the gain drops by 2 dB due to ethanol loss. However, considering the good scattering property of the liquid absorber, this effect is acceptable. The simulated and measured gain of the antenna without ethanol or with ethanol are shown in Fig. 14. The glass glue and background noise are assumed to cause the small difference between simulation and experiment.

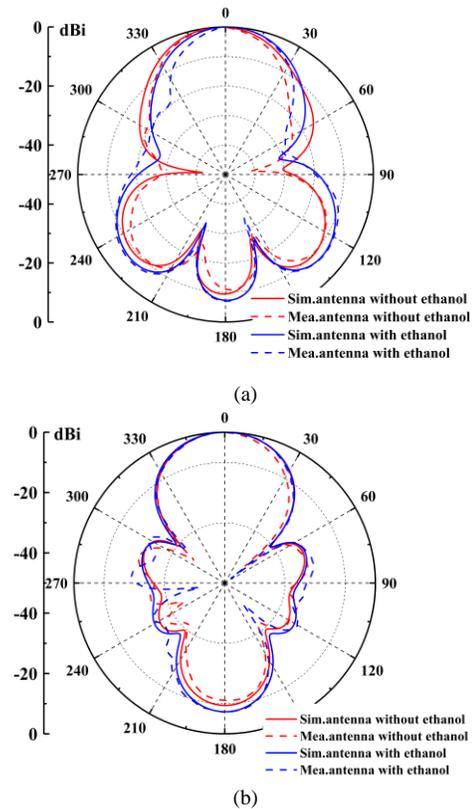

Fig. 12  Simulated and measured radiation patterns of the antenna with ethanol and without ethanol at 3GHz. (a) *xoz*-plane (b) *yoz*-plane.

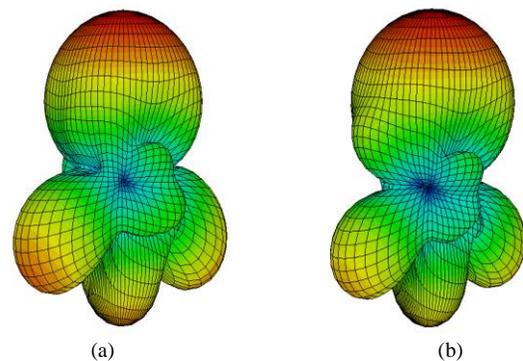

Fig. 13  3-D radiation patterns of the antenna at 3GHz (a) without ethanol and (b) with ethanol.

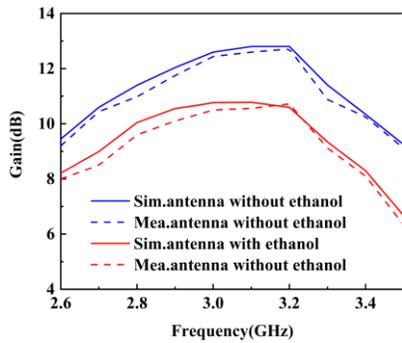

Fig. 14 Simulated and measured gain of the antenna without and with ethanol.

### C. Simulated and Measured Scattering Performance

Fig. 15(a) and (b) shows the monostatic RCS of the antenna with and without ethanol for *x*-polarized and *y*-polarized waves. The measured results are in reasonable agreement with simulations. The following factors contribute to the difference between measurement and experiment: 1) the impinging wave in the simulation is a plane wave, but the radiated wave of the standard horn antenna has a slightly spherical shape; 2) it is difficult to ensure that the container is completely filled with ethanol; 3) background noise in the measurement setup causes measurement tolerance; and 4) a small amount of misalignment is unavoidable in such a setup.

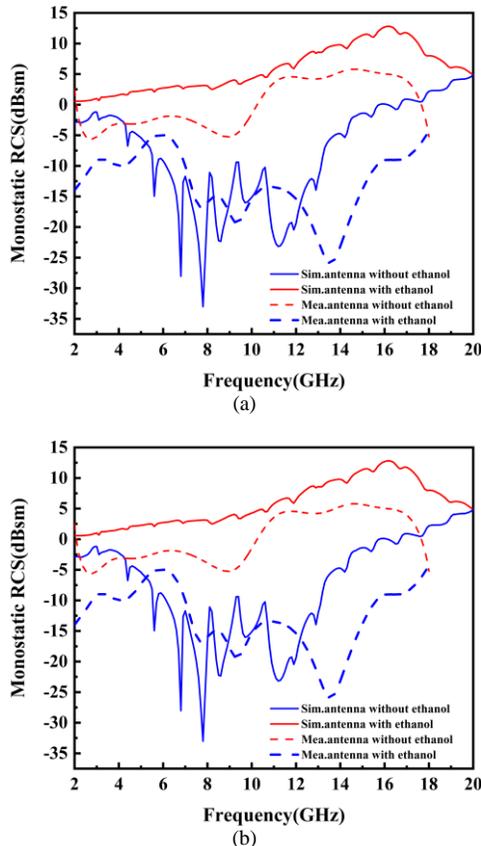

Fig. 15 Simulated and measured monostatic RCS of the antenna with and without ethanol. (a) *x*-polarized wave (b) *y*-polarized wave.

Fig. 16 shows the RCS reduction of the antenna with ethanol compared to the antenna without ethanol. It can be observed that the RCS decreases dramatically within 6 GHz to 16 GHz due to the fact that the absorber absorbs the majority of the power. Also, from 2 GHz to 6 GHz the RCS of the antenna has a small reduction due to the influence of the container and the loss of the ethanol. The resonant frequency points are generated by the resonant cavity formed by the metal reflector and the slot antenna array. The maximum RCS reduction is 36 dB, and the average RCS reduction is 13.28 dB. Within the operating frequency band of the antenna, the bistatic RCS of the antenna without and with ethanol remain almost unchanged, and the slight difference between the two curves is caused by the loss of ethanol, as shown in Fig. 17(a). The bistatic RCS of the antenna is reduced obviously within the operation band of the absorber. This coincides with the operation band of the absorber and indicates that the absorber is useful to reduce the RCS of the antenna. In Fig.17(b) and (c), we can see that the RCS in some directions increases due to the fact that the absorber destroys the resonant pole. Fig. 18 shows that the 3-D patterns of the RCS of the antenna without and with ethanol at 3 GHz, 7.6 GHz and 19.7 GHz. It is found that the absorber reduces the backward RCS while the forward RCS of the antenna remains unchanged due to electromagnetic wave diffraction.

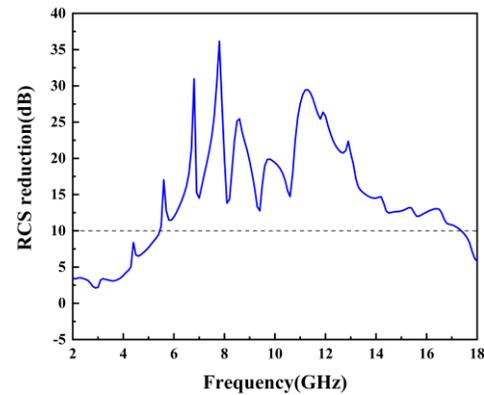

Fig. 16 RCS reduction of the antenna with ethanol compared with the antenna without ethanol.

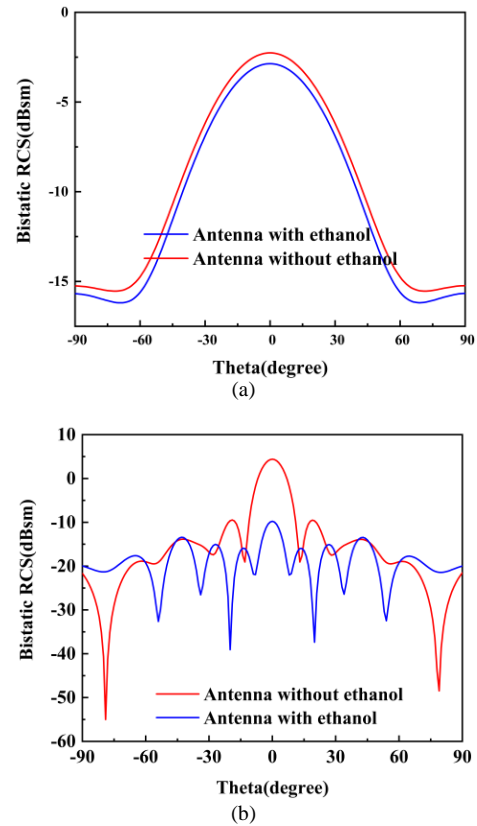

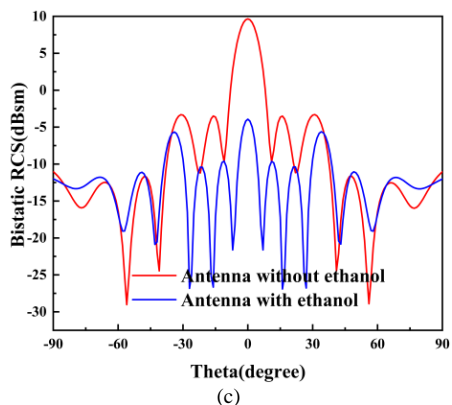

Fig. 17  Bistatic RCS of the antenna without ethanol or with ethanol at  (a) 3 GHz, (b) 7.6 GHz, (c) 19.7GHz.

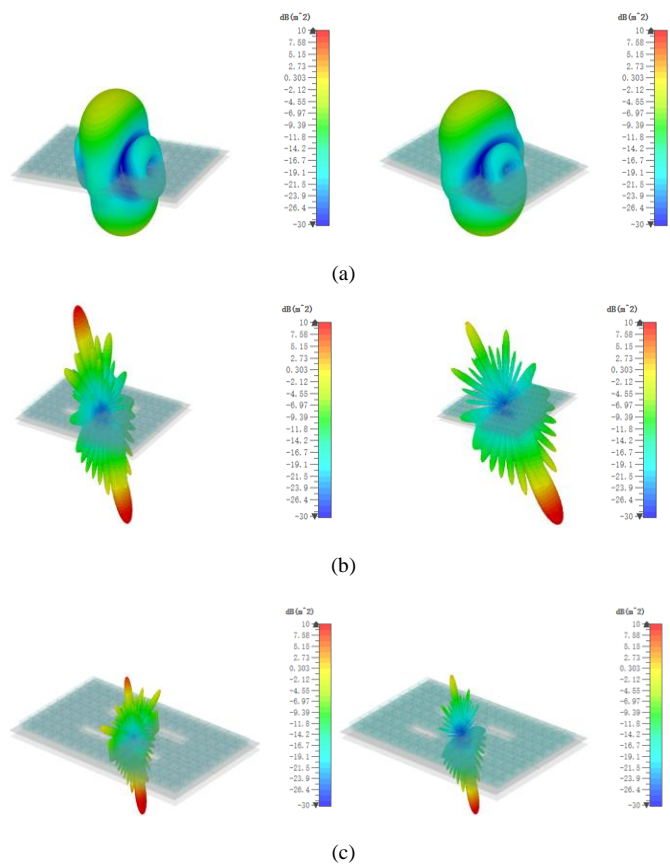

Fig. 18  3-D patterns of RCS of the antenna without ethanol and with ethanol at (a) 3 GHz, (b) 7.6 GHz, (c) 19.7GHz.

## III. Conclusion

This paper proposes a liquid-based reconfigurable RCS slot antenna array. A sandwich-structure liquid absorber with an absorption band ranging from 6.1 GHz to 20.54 GHz is constructed. The 2×2 slot antenna array was combined with the absorber, resulting in wideband monostatic RCS and bistatic RCS reduction. The RCS of the antenna can also be reconfigured by injecting and extracting the ethanol while maintaining the radiation performance of the antenna. The simulated and measured results show that the maximum monostatic RCS of the antenna with ethanol is 36 dB lower than that without ethanol at 6.7 GHz, and the bistatic RCS of the antenna with ethanol is 15 dB lower than that of the antenna without ethanol when $\theta=0°$. The design could be applied in a stealth communication platform in the future.